\documentclass[aps,apl,preprint,endfload*]{revtex4-1}

\usepackage{graphicx}
\usepackage{subfigure}
\usepackage{amssymb}
\usepackage{amsfonts}
\usepackage{amsmath}
\usepackage{amsbsy}

\renewcommand{\vec}[1]{\mbox{\boldmath$\mathrm{#1}$}}

\begin{document}

\title{Thermoelectric effect of multiferroic oxide interfaces}
\author{Chenglong Jia and Jamal Berakdar}
\affiliation{Institut f\"ur Physik, Martin-Luther Universit\"at Halle-Wittenberg, 06099 Halle (Saale), Germany}

\begin{abstract}
   We investigate the thermoelectric  properties of electrons at the interface of oxide heterostructure  and in the presence of a
    multiferroic oxide with spiral spin order. We find there is no (spin) Hall current generated by the temperature gradient.
    A Seebeck effect is however present. Due to the magnetoelectric coupling, the charge and thermal conductivities are electrically controllable via the spin spiral helicity. Moreover,  the thermopower exhibits  a sign change when tuning  electro-statically  the carrier density.
\end{abstract}

%\pacs{}

\maketitle
The efficient manipulation of the spin-dependent 
transport properties of  electronic systems is intensively studied due to its importance in spintronics
applications. Thereby, a key role is played by the  spin-orbit interaction (SOI); e.g.  SOI is at the heart of 
 the charge/spin Hall effect that have been successfully realized in different condensed matter \cite{Nagaosa}. 
  On the other hand,   charge and spin currents can also be generated by a temperature gradient, a phenomena
   termed as the Seebeck and the spin-Seebeck effect, respectively\cite{Book-Marder,SSE}. 
    Very recently,  an intrinsic thermo-spin Hall effect has been predicted in a two-dimensional electron gas (2DEG) with finite Rashba SOI \cite{Ma-10}. 
     A general theory of the thermal Hall effect in quantum magnets is developed by Katsura \emph{et.al.} in Ref.[\onlinecite{KNL}].  The corresponding magnon Hall effect has been observed in Lu$_2$V$_2$O$_7$ with a pyrochlore structure \cite{Magon-HE}. 
     In this letter, we study the thermoelectronics of an interfacial quasi 2DEG formed at the junction 
      of oxides LaTiO$_3$/SrTiO$_3$ or LaAlO$_3$/SrTiO$_3$ \cite{Oxides}. As argued in \cite{JB,HE-JB} new functionalities are achieved
      when one of the oxide layer is 
      multiferroic  with a transverse spiral magnetic order, e.g. RMnO$_3$ (R=Tb, Dy, Gd) \cite{RMnO3}. We expect similar effects
      for compounds of the form   LaAlO$_3$/SrTiO$_3$/RMnO$_3$ when  SrTiO$_3$  is only few layers think. In that case the 
      2DEG  at the interface of LaAlO$_3$/SrTiO$_3$ is expected to be  influenced by the spiral order in RMnO$_3$. The essential point is that 
       due to the topological structure of the local magnetic moments, a traversing carrier  experiences an effective electrically  controllable SOI. This SOI depends linearly  on the carriers wave vector and on the helicity of the oxides magnetic order. Such an effective SOI is in a complete analogy to the semiconductor case where the Rashba\cite{Rashba} SOI and Dresselhaus \cite{Dresselhaus} SOI have equal strengths, and results in an  anisotropic charge and heat conductivities, which is tunable by a transversal electric field, and so is the thermopower.

The non-equlibrium current density in a system subject to potential $V$ and temperature $T$ gradients is phenomenologically given by \cite{Te_Jonson84},
\begin{equation}
\hat{J}^{(i)}_{\mu} = (1/e) \kappa_{\mu \nu}^{(ie)} \partial_{\nu} V + T \kappa_{\mu \nu}^{(iq)} \partial_{\nu} (1/T).
\end{equation}
where $\mu$ and $\nu$ ($=x,y$) are spatial subscripts.  $i$ stands for the subscripts $e$, $q$ and $s$ that denote
respectively  the charge, heat, and spin currents $\hat{\vec J}^{(e)}$, $\hat{\vec J}^{(q)}$, and $\hat{\vec J}^{(s)}$. Those   are defined as
\begin{eqnarray}
  && \hat{\vec v}= \delta H_0 / \delta \vec{P}, \;
   \hat{\vec J}^{(e)} = e \hat{\vec v} \\
  && \hat{\vec J}^{(q)} =\frac{1}{2} \{ H_0, \hat{\vec v} \} - E_f \hat{\vec v}, \\
  && \hat{\vec J}^{(s)} = \frac{\hbar}{4} \{ \vec{\sigma}, \hat{\vec v} \}.
\end{eqnarray}
Here $H_0$ is the unperturbed Hamiltonian, $\vec{P}$ is the momentum operator, $\vec{\sigma}$ is the vector of Pauli matrices, and $E_f$ is the chemical potential of the system. The Seebeck coefficient  $S_{\mu \nu} = - \partial_\mu V /\partial_\nu T$
is given by the zero charge-current condition satisfying
\begin{equation}
  S_{\mu\nu} = (-e/T) \kappa_{\mu \gamma}^{(eq)}/\kappa_{\gamma \nu}^{(ee)}.
\end{equation}
Considering a spatial homogeneous electric field and a temperature gradient, the vector potentials read  $\vec{A}^{(e)} = \vec{E} e^{-i \omega t}/ i\omega$ and $\vec{A}^{(q)} = -\nabla T /T  e^{-i \omega t}/ i\omega$, respectively. Following Luttinger \cite{Luttinger}, all the response functions can be obtained from the Kubo formula \cite{Te_Jonson84,Book_Coleman}
\begin{eqnarray}
  \kappa_{\mu \nu}^{(ij)} (\omega) &=&  \frac{i}{\hbar \omega} \int_0^{\hbar \beta} d\tau
  e^{i\omega \tau} \langle T \hat{J}_{\mu}^{(i)} (\tau) \hat{J}_{\nu}^{(j)} (0) \rangle       \nonumber \\
  &=& \frac{1}{i \omega \beta} \sum_{mn} \langle \psi_m | \hat{J}_{\mu}^{(i)} | \psi_n \rangle \langle \psi_n | \hat{J}_{\nu}^{(j)} |\psi_m \rangle \nonumber \\
  &\times& \sum_p G_{n}(ip+i\omega) G_{m}(ip),
\end{eqnarray}
where $G$ is the Matsubara Green function and only the loop diagram contributing to the current fluctuations is taken into account \cite{Book_Coleman}.
Performing the Matsubara sum over $ip$, we have
\begin{equation}
  \kappa_{\mu \nu}^{(ij)}  (\omega) = \frac{1}{i \omega} \sum_{mn} \frac{[f(E_m) - f(E_n)] (\hat{J}_{\mu}^{(i)})_{mn}  (\hat{J}_{\nu}^{(j)})_{nm} }{E_m - E_n +\omega+ i\Gamma}
\end{equation}
with $\Gamma$ being the relaxation rate.  By using the following relation between the corresponding matrix elements,
\begin{equation}
  \langle \psi_n | \vec{v} |\psi_m \rangle = \frac{i}{\hbar}(E_n - E_m)  \langle \psi_n | \vec{r} |\psi_m \rangle
\end{equation}
the conductivity is obtained in the static limit \cite{Nagaosa,Sinova-04},
\begin{equation}
  \sigma_{\mu \nu}^{(ij)} =\text{Im} \sum_{m\neq n} \frac{f(E_m) - f(E_n)}{(E_m - E_n)/ \hbar }\frac{(\hat{J}_{\mu}^{(i)})_{mn} (\hat{J}_{\nu}^{(j)})_{nm}}{E_m - E_n+ i\Gamma}
\end{equation}

%%%%%%%%%%%%%%%%%%%%%%%%%%%%%%%%%%%%%%%%%%%%%%%%
\begin{figure}[f]
  \centering
  \includegraphics[width=0.6\textwidth]{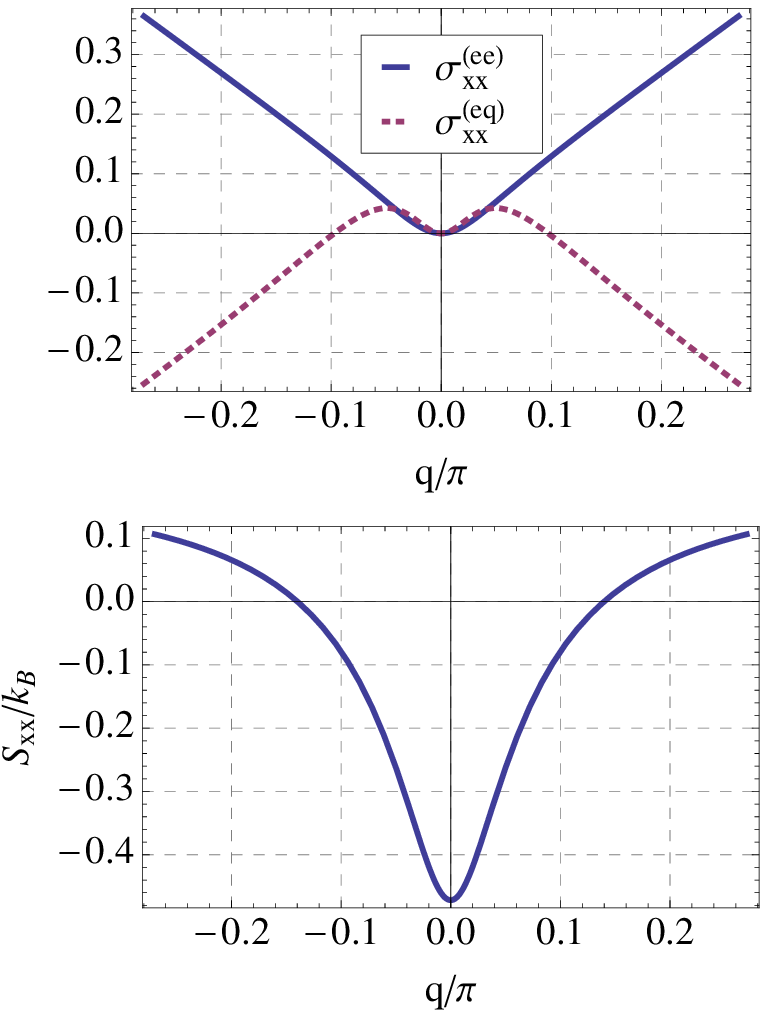}
    \caption{(Top) Charge $\sigma_{xx}^{ee}$[$e^2/2\pi h $] and thermal $\sigma_{xx}^{eq}$ [$(e\hbar/2ma^2)(1/2\pi)^4)$] conductivity. (Bottom) the thermopower as a function of the spiral spin wave vector $q$, where $a$ is the lattice constant and $k_B$ is Boltzmann constant.  
    Parameters are chosen such that $U = 0.1$, $E_f = 0.2$, $\Gamma = 0.05$ and $k_{B}T =0.1$.
    All measured in the unit of energy  $\epsilon_0=\hbar^2/2ma^2$. }
  \label{fig:Sq}
\end{figure}
%%%%%%%%%%%%%%%%%%%%%%%%%%%%%%%%%%%%%%%%%%%%%%%

Our specific system is  the two-dimensional electron gas (2DEG) influenced by  a spiral multiferroic oxide,
as discussed above.  The coupling between the local spiral magnetic moments ($\vec{m}_r$  $ = [ \sin \theta_r, 0, \cos \theta_r ]$
where $\theta_r = \vec{q}_m \cdot \vec{r}$ with $\vec{q}_m = [q,0,0]$)  to the conduction electrons is governed by the $sd$ Hamiltonian \cite{JB},
\begin{equation}
  H=h_k+h_{sd}=\frac{1}{2m}\vec{P}^2 + U \vec{n}_r \cdot \vec{\sigma}.
\end{equation}
After applying an unitary local gauge transformation in the spin space
$U_g = e^{-i {\theta_r} \sigma_y/2}$, we get
\begin{equation}
  H = \frac{1}{2m}(\vec{P} + \vec{A}_g)^2 + U \tilde{\sigma}_z
  \label{eq:h}
\end{equation}
where $\vec{A}_g = -i \hbar U_g^{\dagger} \nabla_{\vec{r}} U_g = ( -\hbar q \tilde{\sigma}_y/2) \hat{e}_x $ is the topological vector potential (hereafter, transformed quantities have a tilde).
Explicitly diagonalizing the  Hamiltonian (\ref{eq:h}) we obtain the eigenenergies
\begin{equation}
  E_{\pm}(\vec{k}) = \frac{\hbar^2 \vec{k}^2}{2m} \pm \sqrt{U^2 + (\frac{\hbar^2 q k_x}{2m})^2}
  \label{eigenenergies}
\end{equation}
with the eigenstates
\begin{eqnarray}
\label{eigenstates}
  |\psi_{+} \rangle = e^{-i \vec{k} \cdot  \vec{r}} \left( \begin{array}{cc}  \cos  \frac{\phi}{2} \\ i \sin \frac{\phi}{2} \end{array} \right),\;
  |\psi_{-} \rangle = e^{-i \vec{k} \cdot  \vec{r}}  \left( \begin{array}{cc}  i \sin  \frac{\phi}{2} \\ \cos \frac{\phi}{2} \end{array} \right) \nonumber
\end{eqnarray}
where
\begin{equation}
  \tan \phi=  \frac{\hbar^2 qk_x}{2mU},\; \; \cos \phi = \frac{U}{\sqrt{U^2 + (\frac{\hbar^2 q k_x}{2m})^2}}.
\end{equation}
The velocity operators then read,
\begin{eqnarray}
  \hat{v}_x = -i \frac{\hbar}{m}\overleftrightarrow{\nabla}_x -\frac{\hbar q\tilde{\sigma}_y}{2m},\; \hat{v}_y = -i \frac{\hbar}{m}\overleftrightarrow{\nabla}_y
\end{eqnarray}
where $\overleftrightarrow{\nabla} = (\overrightarrow{\nabla} -\overleftarrow{\nabla})/2$ is the symmetrized derivative. The topological vector potential $\vec{A}_g$ introduces an extra \emph{dynamical diamagnetic} response (the last term of $\hat{v}_x$) to all currents. However,  different from the case of 2DEG embedded in a magnetic filed \cite{Te_Jonson84}, the corresponding dia-thermal current has zero equilibrium expectation value since $\langle \tilde{\sigma}_y \rangle =0$,  whereas there is a persistent dia-spin current $\langle \hat{J}^{(s)}_x \rangle$ along the the spin wave vector of the spiral \cite{JB}.  Both the charge ($\langle \hat{J}^{(e)}_{\mu} \rangle$) and the heat ($\langle \hat{J}^{(q)}_{\mu} \rangle$) currents are generated along the direction of the external electric field and temperature gradient.  We have just the Seebeck effect in absence of the spin-Seebeck effect.

%%%%%%%%%%%%%%%%%%%%%%%%%%%%%%%%%%%%%%%%%%%%%%%%
\begin{figure}[f]
  \centering
  \includegraphics[width=0.6\textwidth]{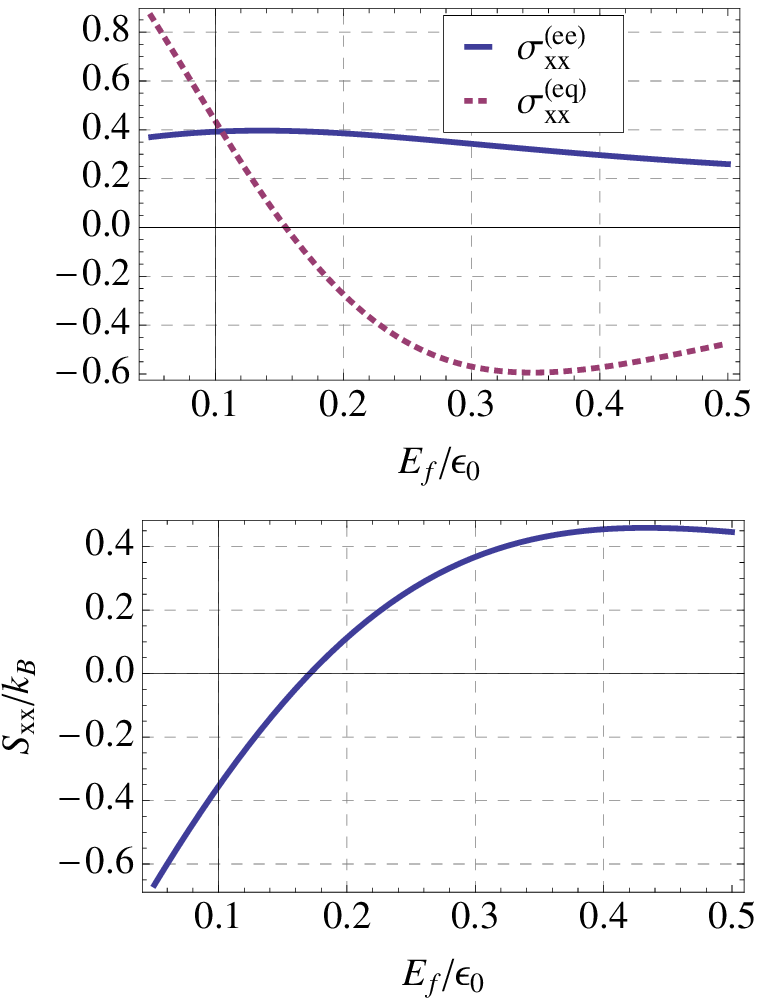}
  \caption{(Top) Charge $\sigma_{xx}^{ee}$[$e^2/2\pi h $] and thermal $\sigma_{xx}^{eq}$[$(e\hbar/2ma^2)(1/2\pi)^4]$ conductivity and (bottom) the Seebeck coefficient $S_{xx}$ vs. chemical potential $E_f$ with the spiral helicity $q =2\pi/7$. Other parameters are same as in Fig.\ref{fig:Sq}.}
  \label{fig:Su}
\end{figure}
%%%%%%%%%%%%%%%%%%%%%%%%%%%%%%%%%%%%%%%%%%%%%%%

Now let's consider the transport properties. It is straightforward to work out that the only nonzero matrix elements are

\begin{eqnarray}
 && \langle \psi_{+} | \hat{J}_x^{(e)} | \psi_{-} \rangle = ie \frac{\hbar q}{2m} \cos\phi, \\
 && \langle \psi_{+} | \hat{J}_x^{(q)} | \psi_{-} \rangle = i\frac{\hbar q}{2m} (\frac{\hbar^2 \vec{k}^2}{2m} - E_f) \cos\phi, \\
 && \langle \psi_{+} | \hat{J}_{\mu}^{(s)} | \psi_{-} \rangle = i\frac{\hbar^2 k_{\mu}}{2m} \cos\phi .
\end{eqnarray}
which results in
\begin{equation}
  \sigma_{\mu \nu }^{(sj)} \equiv 0 \; \text{and } \sigma_{y \mu}^{(eq/qe)} \equiv 0.
\end{equation}
No  Hall current is induced by the external electric field nor by the temperature gradient because of the 
 disappearance of the Berry phase \cite{Nagaosa, Sinova-04, Shen-04} due to  the resonant form of the eigenstates as a results of   the coplanar spiral magnetic moments \cite{JB}. The charge and the heat conductivity is anisotropic, only the diagonal component of the conductivities is nonzero. That is is sharp contrast to the semiconductor 2DEG with equal Rashba and Dresselhaus SOI leading to an isotropic conductivity \cite{MH-03, MBM-06}.  The non-vanishing components of the conductivity tensors are presented in Fig.\ref{fig:Sq}.  It should be noted that, due to the 
  presence of the  magneto-electric coupling, all transport properties are tunable by a small transverse electric field ($\sim$ 1kV/cm)  which tunes the spin helicity $q$  \cite{E-Control-1}.  For a large electric field, the FE polarization is stabilized, but the concentration of carriers at the interface is modulated \cite{E-Control-2}. We thus have an electrically controlled chemical potential, which results in a sign change of the thermal conductivity and of the Seebeck coefficient (see Fig.\ref{fig:Su}).

When the oxide magnetic moments possess a small deviation from the spiral plane,  the scalar spin chirality defined as the mixed product of three spins on a certain plaquette, $\chi_{ijk}=\vec{S}_i \cdot (\vec{S}_j \times \vec{S}_k)$ becomes nonzero. $\chi_{ijk}$ introduces a fictitious magnetic flux to the conduction electrons and provides a nontrivial Berry curvature of the wave function, leading to nonzero charge/spin \cite{JB,Kagome} and thermal \cite{KNL} Hall conductivity.

\newpage
%%%%%%%%%%%%%%%%%%%%%%%%%%%%%%%%%%%%%%%%%%%%%%%

%
\end{document}